# Aza-Triangulene: On-Surface Synthesis, Electronic and Magnetic Properties

Tao Wang,[1,2,‡,*] Alejandro Berdonces-Layunta,[1,2,‡] Niklas Friedrich,[3] Manuel Vilas-Varela,[4] Jan Patrick Calupitan,[2,*] Jose Ignacio Pascual,[3,5] Diego Peña,[4] David Casanova,[1,5] Martina Corso,[1,2] and Dimas G. de Oteyza[1,2,5,6,*]

[1]Donostia International Physics Center, 20018 San Sebastián, Spain

[2]Centro de Fisica de Materiales CFM/MPC, CSIC-UPV/EHU, 20018 San Sebastián, Spain

[3]CIC nanoGUNE BRTA, 20018 San Sebastián, Spain

[4]Centro Singular de Investigación en Química Biolóxica e Materiais Moleculares (CiQUS) y Departamento de Química Orgánica, Universidade de Santiago de Compostela, 15782 Santiago de Compostela, Spain

[5]Ikerbasque, Basque Foundation for Science, 48009 Bilbao, Spain

[6]Nanomaterials and Nanotechnology Research Center (CINN), CSIC-UNIOVI-PA; 33940 El Entrego, Spain

**ABSTRACT:** Nitrogen heteroatom doping into a triangulene molecule allows tuning its magnetic state. However, the synthesis of the nitrogen-doped triangulene (aza-triangulene) has been challenging. Herein, we report the successful synthesis of aza-triangulene on the Au(111) and Ag(111) surfaces, along with their characterizations by scanning tunneling microscopy and spectroscopy in combination with density functional theory (DFT) calculations. Aza-triangulenes were obtained by reducing ketone-substituted precursors. Exposure to atomic hydrogen followed by thermal annealing and, when necessary, manipulations with the scanning probe, afforded the target product. We demonstrate that on Au(111) aza-triangulene donates an electron to the substrate and exhibits an open-shell triplet ground state. This is derived from the different Kondo resonances of the final aza-triangulene product and a series of intermediates on Au(111). Experimentally mapped molecular orbitals match perfectly with DFT calculated counterparts for a positively charged aza-triangulene. In contrast, aza-triangulene on Ag(111) receives an extra electron from the substrate and displays a closed-shell character. Our study reveals the electronic properties of aza-triangulene on different metal surfaces and offers an efficient approach for the fabrication of new hydrocarbon structures, including reactive open-shell molecules.

Triangulene, as the smallest triplet-ground-state polybenzenoid, has attracted intensive attention since it was theoretically devised back in 1953.[1] In spite of its even number of carbon atoms, it is not possible to pair up all of its π-electrons to form a closed-shell structure.[2–4] The total net spin of triangulene in its ground state is quantified by Ovchinnikov's rule[5] and Lieb's theorem[6] for bipartite lattices: $S=(N_A-N_B)/2$, where $N_A$ and $N_B$ denote the numbers of carbon atoms belonging to each of the two sublattices ($N_A=12$, $N_B=10$, $S=1$, Figure 1a). Due to its high reactivity stemming from its unpaired electrons, the synthesis of triangulene by conventional solution-phase chemistry has been inaccessible.[1,4] Fortunately, the recently developed on-surface synthesis (OSS)[7–10] under ultra-high vacuum (UHV) conditions opens a door for the fabrication of reactive carbon-based structures holding π-radicals, where a rationally designed precursor is annealed at high temperatures over a catalytic surface in order to form the target product.[11–13] Using the OSS strategy, triangulene and other extended [n]triangulenes (n>3) have been successfully synthesized, whose structures were characterized precisely with the aid of bond-resolving scanning tunneling microscopy (BR-STM) and non-contact atomic force microscopy (nc-AFM) techniques.[3,14–17] In addition, the open-shell character of triangulene and its derivates was confirmed by the observation of singly occupied/unoccupied molecular orbitals (SOMO/SUMO),[14,15] Kondo resonances,[18] and spin-flip excitations.[17–19]

Heteroatom doping can substantially modify the electronic and magnetic properties of graphene-based structures.[20–26] For example, the substitution of the central carbon atom of triangulene by a nitrogen adds a π-electron to the system, resulting in different groundstate spin multiplicity with respect to that of undoped triangulene. In a naïve picture, one may assume a double occupancy of the $p_z$ orbital on the central nitrogen atom of aza-triangulene. This would remove its contribution to the $N_B$ count in Ovchinnikov's rule and result in a quartet ground state ($N_A=12$, $N_B=9$, $S=3/2$, Figure 1a). The same may be expected from the chemical structure drawing as shown in structure **7** (Figure 1b), representing a molecule with $D_{3h}$ symmetry and three radicals, one on each molecular side. All are located on the same sublattice and are thus expected to align ferromagnetically. However, theoretical calculations predict that for aza-triangulene,[27] a doublet ground state ($S=1/2$) with a $C_{2v}$ molecular symmetry, driven by a Jahn–Teller distortion,[28] is energetically more favorable than the more intuitive $D_{3h}$ structure.

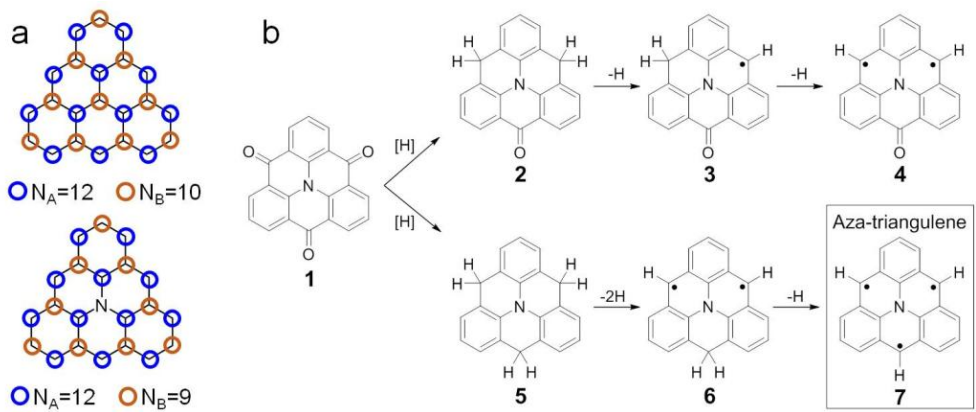

**Figure 1.** (a) Schematic representation of carbon atoms belonging to each of the two sublattices. (b) Reaction steps for the synthesis of aza-triangulene **7** on Au(111), starting from the hydrogenation of precursor **1**, followed by 250 °C annealing and subsequent tip manipulation. The final product, intermediates, and byproducts observed in experiments are presented.

Despite these interesting theoretical predictions,[27] synthesis of aza-triangulene had, until now, remained elusive. Herein, we report the on-surface synthesis of aza-triangulene on both Au(111) and Ag(111). Figure 1b shows the reaction procedure starting from the corresponding precursor, that is, the ketone-substituted aza-triangulene **1**. In our previous work, the combination of atomic hydrogen reduction followed by annealing was shown efficient in removing the oxygen atoms on ketone-substituted graphene nanoribbons on Au(111).[29] Inspired by this, we employed a similar procedure for molecule **1**. Hydrogen reduction followed by annealing resulted in partial (**2**) or complete (**5**) deoxygenation of the precursor **1**. Subsequent tip-induced removal of hydrogen atoms results in byproducts **3** and **4**, intermediate **6**, and the final product aza-triangulene **7** (Figure 1b). We show that aza-triangulene donates an electron to the Au(111) substrate and bears a triplet ground state. On the contrary, aza-triangulene receives an electron from the low-work function Ag(111) surface, resulting in a closed-shell ground state.

**Synthesis.** Precursor molecule **1** was obtained by solution-phase synthesis following a previously reported procedure (Figure S1; Supporting information).[30,31] Figure 2a shows the STM image of the sample upon depositing **1** on Au(111) held at room temperature (RT). The molecules self-assemble into well-ordered dense islands stabilized by a two-dimensional network of intermolecular hydrogen bonds. BR-STM imaging (Figure 2a, inset) shows the ketone groups exhibiting V-shape protrusions pointing toward the hydrogen atoms of the neighboring molecules. The energy and spatial distribution of frontier molecular orbitals as obtained with scanning tunneling spectroscopy show an excellent agreement with previous reports (Figure S2).[30]

The sample held at RT was then hydrogenated by exposing to atomic hydrogen. As a result, a considerable number of $sp^3$-hybridized carbon atoms were generated (Figure S3), which turned most of the molecules three-dimensional.[29] Subsequent annealing at 250 °C caused a notable desorption and planarized most of the remaining molecular species (>70 %; Figure 2b). Figure 2c shows a representative constant height STM image of the most abundant product in Figure 2b. The removal of a hydrogen from a $CH_2$ edge atom requires an annealing temperature around 300 °C,[34,35] higher than that used to dehydrogenate the "interior" carbon atoms. Our thermal treatment therefore still maintains $sp^3$-hybridized carbon atoms at the molecular edges that weaken molecule-substrate interactions,[32] thus facilitating molecular diffusion and hindering the acquisition of BR-STM images. However, since hydrogen atoms tend to passivate the most reactive sites of nanographenes[12,33] (which in this case are the center carbon atoms at the three zigzag edges of aza-triangulene), structure **5** is the most probable major product.

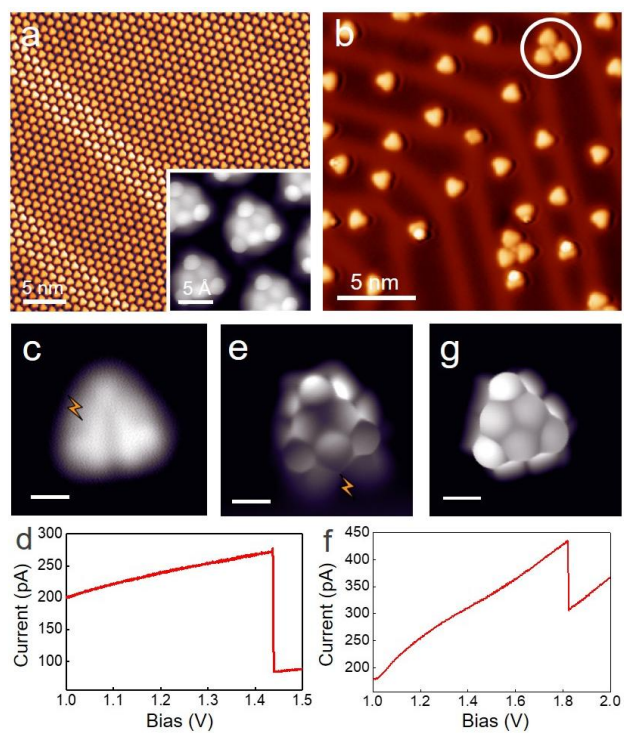

**Figure 2.** (a) STM image of the sample prepared by depositing molecule **1** on Au(111) at RT. Inset shows a BR-STM image taken by a CO-functionalized probe.[36] (b) STM image of the sample after hydrogenation and annealing at 250 °C. A self-assembled triangulene trimer is marked with a white circle. (c,e,g) Constant-height STM images of molecules **5**, **6**, **7**, respectively. (d,f) I-V spectroscopy demonstrating the tip-induced removal of additional hydrogens. The positions for the tip induced pulses are marked in (c) and (e). Tunneling parameters: (a,b) U=−1 V, I=−100 pA; inset of (a) and (c,e,g): U=5 mV. Scale bars in (c, e, g) are 5 Å.

Tip induced dehydrogenation was then employed to remove H atoms from the $sp^3$-hybridized carbons of **5** (Figure 2c-f). We ramped up the bias with the tip positioned over the zigzag edge of this product (marked as a lightning in Figure 2c). The dehydrogenation is observed as a current step in I-V spectra, as shown in Figure 2d. Subsequent BR-STM images allow for assignment of the intermediate product structure as **6** (Figure 2e). The six-membered ring containing the remaining $sp^3$ carbon atom is much larger than others and exhibits a sharp corner—a widely reported fingerprint of $sp^3$-carbon containing rings.[3,12,33,34] In all our attempts, the first dehydrogenation step occurred directly from **5** to **6**, *i.e.* by simultaneously removing two H atoms from two edges. The final product **7** was generated by the tip-induced removal of the last residual hydrogen from **6**, as confirmed by the BR-STM image in Figure 2g. It is worth mentioning that, apart from the tip manipulation approach, product **7** can also be produced directly by annealing the sample shown in Figure 2b at a temperature above 300 °C. However, because the spin density of open-shell molecules automatically boosts their reactivity,[37] most of the molecules react and appear as dimers or oligomers. An example is presented in Figure S4, in which **7** and various products from molecular fusions coexist.

As marked by the white circle in Figure 2b, a few trimers with a dot in the center are also observed on this sample. These molecules retain a residual ketone group (**2**; Figure 1) and the trimer structure is apparently stabilized *via* O···Au coordination interactions and hydrogen bonds (see details in Figure S5-S7), as reported previously for other ketone-functionalized carbon-nanostructures.[29,38] Using similar tip manipulation procedures as described above, products **3** and **4** (Figure 1b) can be obtained hierarchically as well (Figure S5-S6).

**Kondo resonance and charge transfer.** Next, we investigate the magnetism of **3**, **4**, **6**, and **7**, which are all expected to be open-shell systems as predicted by DFT calculations in vacuum (Figure 3b). Figure 3a shows their corresponding low-energy dI/dV spectra on Au(111). The spectrum of **3** does not exhibit any visible signal (apart from the two well-known inelastic vibrational modes of the CO molecule at the tip apex at ~5 mV and ~35 mV), implying a closed-shell structure.[39,40] The spectra from **4**, **6**, and **7** all exhibit symmetric zero-energy peaks around the Fermi level that are attributed to Kondo resonances,[41] as widely reported in metal-supported open-shell carbon structures.[11] This is further supported by the temperature-dependent spectra of intermediate **6** (Figure S8). Fitting the spectra with a Frota function[42] reveals a rapid peak broadening with increasing temperature, following the characteristic trend of Kondo resonances,[12,13,43] which originates from the screening of the local spin by the conduction electrons of the underlying metal substrate.[41,44] The Kondo resonance from **4** and **6** have high and comparable amplitudes and also similar FWHMs (full width at half maximum; 8.1±0.6 mV for **4** and 8.6±0.1 for **6**; derived from six and two data points respectively) while the Kondo resonance of **7** on Au(111) is much weaker and has an apparently larger FWHM (13.0±2 mV; derived from six data points). This hints at **4** and **6** probably displaying a doublet ground state (S=1/2) whereas **7** presumably holds a high-spin ground state (S⩾1), whose underscreened Kondo peaks typically display much lower amplitude than those from a S=1/2 system.[18,33,45]

The proposed ground-state spin for the four molecules (**3**, **4**, **6**, **7**) based on the registered Kondo resonances do not match the DFT predictions for the neutral molecules (Figure 3b). However, spin multiplicities computed for the corresponding cationic species **3**⁺ (S=0, closed-shell), **4**⁺ and **6**⁺ (S=1/2), and **7**⁺ (S=1) are in excellent agreement with the experimental characterization (Figure 3c). In other words, electron transfer from the molecule to the Au(111) substrate reconciles the experiments with the theory.

The charge transfer can be understood from the high work function of the Au(111) surface and the associated low binding energy of highest occupied molecular orbitals (HOMO) levels of hydrocarbon structures atop,[46] along with the n-doping effect of graphitic N-subsituents.[47] Indeed, similar charge transfer processes were observed in a number of N-doped molecules on Au(111).[48–50] In contrast, no charge transfer was detected in the previous works studying unsubstituted extended triangulenes on Au(111).[14,15] These findings agree with the lower ionization energy of aza-triangulene (5.0 eV, see Figure S9) as compared to those of unsubstituted triangulene and extended triangulenes (6.3 eV and 6.2 eV).

**Electronic properties of aza-triangulene on surfaces.** In the following, we focus our attention to the electronic structure around the Fermi level of aza-triangulene (Figure 4). On Au(111), the dI/dV spectroscopy on **7** presents three prominent peaks, at −1.35, 0.35 and 1.2 V (Figure 4a). The spatial maps of the dI/dV signal at −1.35 and 0.35 V are identical (Figure 4c, d) and can be associated to the SOMOs and SUMOs of **7**⁺, respectively, separated by a 1.7 eV Coulomb gap.[13] In fact, also the amplitude of the Kondo resonance appears with the same distribution as the SOMOs and SUMOs (Figure S10), supporting that the Kondo resonance originates from the SOMOs of the cationic aza-triangulene. In turn, the resonance at 1.2 V corresponds to the lowest unoccupied molecular orbital (LUMO; Figure 4b). The experimental dI/dV maps of all these orbitals exhibit three-fold symmetry and match well with the corresponding DFT-calculated density of states (DOS) for the positively charged aza-triangulene (Figure 4e; more details in Figure S11). In contrast, the conductance maps do not fit the DOS distribution of the frontier orbitals of the neutral aza-triangulene (Figure S11), corresponding to non-degenerate irreducible representations of the $C_{2v}$ symmetry point group (Figure 4f and Figure S11).

It is known that the energy level alignment in molecule-substrate dyads hinge on the substrate's work function. Ag(111) displays a work function that is ~0.6 eV lower than that of Au(111) (~4.7 *vs.* ~5.3 eV),[51] a difference that may be sufficient to prevent the electron transfer from the molecule to the substrate. Therefore, we used Ag(111) as substrate to produce aza-triangulene by the same hydrogenation procedure as on Au(111), followed by a 300 °C annealing treatment (see Figure S12). Neither a Kondo resonance nor any spin excitation were detected for aza-triangulene on Ag(111) (Figure S12), suggesting a closed-shell ground state. This is clearly indicative that, rather than remaining as a neutral molecule with a doublet ground state, one electron was transferred from Ag(111) to aza-triangulene.[52]

According to the DFT-calculated molecular orbitals (Figure 4g), aza-triangulene indeed becomes closed-shell when it receives an extra electron and, like **7**⁺, **7**⁻ recovers three-fold ($D_{3h}$) symmetry since the doubly degenerate (e'') HOMOs are fully occupied and Jahn-Teller distortions are deactivated.

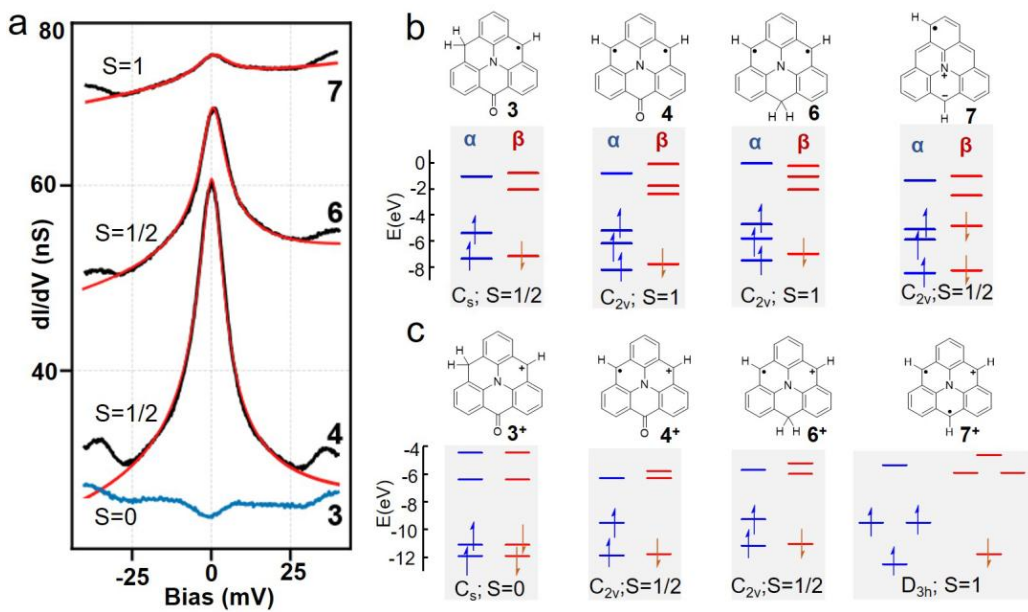

**Figure 3.** (a) Low-energy dI/dV spectra taken of molecules **3**, **4**, **6**, and **7**, on Au(111). Lock-in amplitude: 2 mV. A Frota function is used to fit the spectra from **4**, **6**, **7**, which are attributed to Kondo resonances. (b, c) DFT calculated energy levels of frontier molecular α and β spin-orbitals of the (b) neutral and (c) positively charged molecules **3**, **4**, **6**, and **7** (in vacuum). The drawn resonance structure of **7** will be further discussed in Figure 5 and its corresponding text.

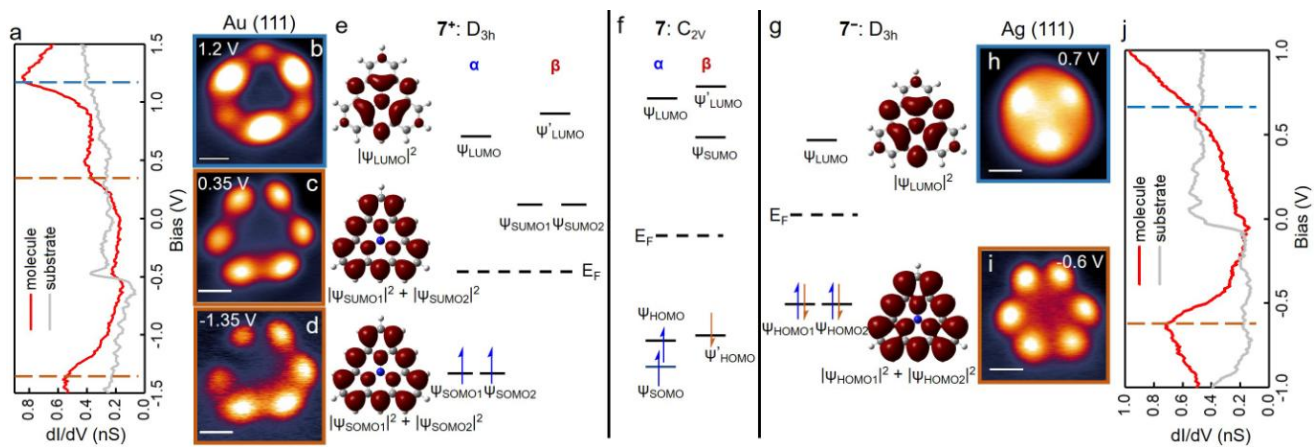

**Figure 4.** (a, g) Long-range dI/dV spectra taken on aza-triangulene on Au(111) and Ag(111), respectively. Lock-in amplitude: 20 mV. (b-d) Constant-current dI/dV maps of aza-triangulene on Au(111) using a metal tip, at the energies of 1.2, 0.35, and -1.35 V, respectively. (e,g) DFT calculated spatial distribution of electron density of states corresponding to frontier molecular orbitals of aza-triangulene on Au(111) and Ag(111), respectively. The calculated electronic density of states is from the summation of the squares of the calculated degenerate orbitals. (f) Energy levels of the neutral N-trangulene. (h,i) Constant-current dI/dV maps of aza-triangulene on Au(111) using a metal tip, at the energies of 0.7 and -0.6 V, respectively. All the scale bars are 5 Å.

In the long-range dI/dV spectra of aza-triangulene on Ag(111) (Figure 4j), two electronic resonances are clearly resolved at -0.6 and 0.7 V bias, which we assign to HOMOs and LUMO, respectively. dI/dV maps at these two energies (Figure 4g, h) exhibit three-fold symmetry and match nicely with the DFT-calculated spatial distribution of the HOMOs and LUMO DOS of **7⁻** (Figure 4h, i).

**Discussion.** Based on above observations, charge transfer between aza-triangulene and the substrate increases the molecular symmetry (from $C_{2v}$ to $D_{3h}$). Whereas the neutral molecule is predicted to display a $C_{2v}$ symmetry in its ground state, DFT calculations disclose, in agreement with experiments, that positively and negatively charged aza-triangulenes exhibit a $D_{3h}$ symmetry instead. Moreover, while **7⁺** holds a ferromagnetic ground state (S=1), **7⁻** is a closed-shell (S=0) species.

In the following, we further rationalize the change of symmetry upon charge transfer (Figure 5). According to calculations,[27] a neutral aza-triangulene with $D_{3h}$ symmetry (S=3/2) has longer carbon-nitrogen bond lengths (Figure 5a) than in the $C_{2v}$ conformation (Figure 5b). This suggests a lower bond order for the C-N bonds in the $D_{3h}$ configuration, resulting in a resonant structure with three unpaired π radicals (Figure 5a) delocalized evenly around the three-fold symmetric molecule and the nitrogen $p_z$ orbital not participating in the conjugated molecular π-network. Note that, for this structure, the calculated spin density on N (Figure 5c) is parallel to that of the three neighboring C atoms, which goes against the antiferromagnetic alignment of electronic spins in chemical bonds and is thus in agreement with the absence of a C–N π-bond.[53] In contrast, adoption of a $C_{2v}$ symmetry with S=1/2 (Figure 5b), driven by a Jahn-Teller distortion, reduces the C–N bond-lengths (lowest with one particular neighboring carbon atom; Figure 5b). A resonance structure that may correspond to this configuration involves a zwitterionic structure and a C–N π-bond (Figure 5b). This is also supported with the calculated spin density, which reveals no spin frustration,[53] *i.e.*, antiferromagnetic spin polarization interactions for all neighboring atoms (Figure 5d), stabilizing the $C_{2v}$ spin doublet (S=1/2) structure by 0.49 eV with respect to the $D_{3h}$ quartet.[27] Besides, the agreement between the calculated spin density distribution and the proposed location of the π radical also matches the suggested zwitterionic structure (Figure 5b, d). Lastly, it also reconciles the ground state spin 1/2 with Ovchinnikov's rule, since the bonding nature of the N $p_z$ orbital justifies its counting towards $N_B$, whereas the carbon atom hosting the negative charge at the low side edge has its $p_z$ orbital doubly occupied and does not count towards $N_A$ ($N_A$=11, $N_B$=10, S=1/2).

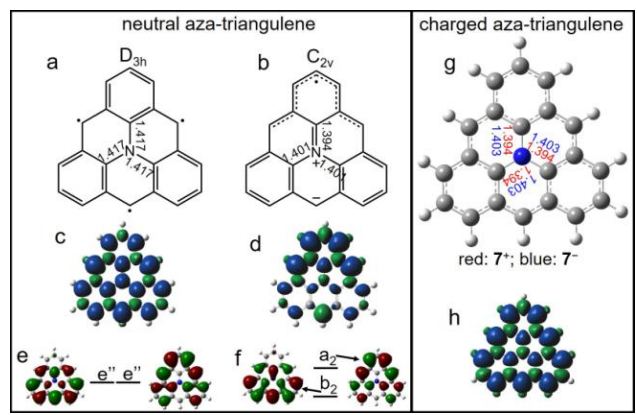

**Figure 5.** (a,b) Molecular structure of neutral aza-triangulene with $D_{3h}$ and $C_{2v}$ symmetry, respectively, along with the calculated length of the carbon-nitrogen bonds (Å). (c,d) Spin density distributions of neutral aza-triangulene with $D_{3h}$ and $C_{2v}$ symmetry, respectively. (e,f) DFT frontier molecular orbitals of **7** with $D_{3h}$ and $C_{2v}$ symmetry, respectively, along with their relative energy alignments. (g) Molecular model of charged aza-triangulenes. C-N bond lengths are indicated as red and blue numbers (Å) for positively and negatively charged cases, respectively. (h) Spin density distribution of $7^+$ (S=1).

However, whereas the picture above rationalizes the symmetry and net spin for the neutral species **7**, it does not explain the symmetry change upon molecular charging. This can be understood from the distinct filling of the frontier orbitals. In the $D_{3h}$ symmetry, the two-fold e'' orbitals present larger electron density probability at the molecular edges (Figure 5e), with a three-fold symmetric combined DOS. Neutral aza-triangulene has an odd number of $p_z$ electrons with three electrons in the e'' orbital space, which prevents an equal orbital occupation and induces a Jahn-Teller distortion towards the $C_{2v}$ structure (Figure 5b) lowering the overall energy of the system. Frontier orbitals of the $C_{2v}$ structure ($a_2$ and $b_2$; Figure 5f) present spatial distributions similar to the e'' pair in the $D_{3h}$ arrangement, with two electrons filling the $b_2$ orbital and one in the $a_2$ orbital localized on the molecular vertex crossed by the $C_2$ axis. As a result, the two-fold symmetric vertex of the molecule remains with a lower electronic density but higher spin density (Figure 5d).[54]

Positively and negatively charged aza-triangulene molecules ($7^+$ and $7^-$) hold an even number of $p_z$ electrons, allowing for an equal population of the e'' frontier orbital pair. In $7^+$ the e'' space is partially occupied and vibronic couplings inducing $D_{3h} \rightarrow C_{2v}$ Jahn-Teller distortion, as in neutral **7**, compete with Hund's-rule coupling emerging from Coulomb repulsion and exchange interaction of the two electrons in the two e'' orbitals. In this case, the Jahn-Teller induced orbital stabilization is not enough to overcome Hund's rule and, as a result, $7^+$ has a spin-triplet $D_{3h}$ ground state (like its isoelectronic sister molecule pristine triangulene)[3] with the positive charge delocalized over the whole molecule. The calculated structure reveals a reduced C-N bond length comparable to that of the neutral $C_{2v}$ structure, hinting at the involvement of the N-atom in the π-conjugation network, *i.e.,* higher bond order and π-character of the C-N bond (Figure 5g). Indeed, this is also supported by the calculated spin density (Figure 5h), which, in contrast to the neutral $D_{3h}$ case (Figure 5c), now shows antiferromagnetic spin orientations between the N and its neighboring C atoms, as well as between all neighboring atoms throughout the whole molecule (Figure 5h).

In $7^-$, the e'' orbitals become both doubly occupied, resulting in a closed-shell structure with three-fold symmetry and no net spin. Also here the calculated bond length for C-N (Figure 5g) is shorter than for the neutral $D_{3h}$ structure, suggesting again the involvement of the N atom in the π-conjugation network.

Altogether, this is a simplistic chemical view of symmetries *vis-à-vis* the charge state of the molecule: the neutral aza-triangulene adopts a conformation, *i.e.*, symmetry and resonance structure, that minimizes the number of unpaired spins at the expense of lowering its symmetry and generating localized charges in the zwitterionic form. In turn, the charged molecules adopt a three-fold symmetric conformation that delocalizes the charge and leads to closed-shell ($7^-$) or open-shell ($7^+$) structures.

**Conclusion.** In summary, we synthesized nitrogen-doped triangulene on both Au(111) and Ag(111) surfaces by H reduction followed by annealing and tip manipulations. The Kondo resonances of different intermediates and products provide evidences that charge transfer sets in from molecules to the Au(111) substrate and the aza-triangulene acquires a triplet open-shell ground state. This is further confirmed by the excellent agreement between the experimentally obtained conductance maps at the energies of the frontier molecular orbitals of aza-triangulene and DFT calculated counterparts on the positively charged aza-triangulene. Opposite from the case on Au(111), a low-work-function Ag(111) substrate donates an electron to aza-triangulene, leading to the formation of a closed-shell structure. In addition, we provide a chemically intuitive picture for the origin of the $C_{2v}$ symmetry of a neutral aza-triangulene molecule, and rationalize the symmetry divergences of its charged states.

## ASSOCIATED CONTENT

**Supporting Information**. Methods, additional STM images and more details of DFT calculations. This material is available free of charge via the Internet at http://pubs.acs.org.

Corresponding Author

* taowang@dipc.org; janpatrick.calupitan@ehu.eus; d_g_oteyza@ehu.eus.

Author Contributions

The manuscript was written through contributions of all authors. / All authors have given approval to the final version of the manuscript. / ‡These authors contributed equally.

Notes

The authors declare no competing financial interest.

ACKNOWLEDGMENT

We acknowledge financial support from MCIN/AEI/10.13039/501100011033 (grant nos. PID2019-107338RB-C61, PID2019-107338RB-C62, PID2019-107338RB-C63, PID2019-109555GB-I00 and FJC2019-041202-I); the Basque Government (IT-1255-19 and PIBA19-0004); and the Spanish Research Council (ILINKC20002), the European Union's Horizon 2020 research and innovation program (grant no. 863098 and Marie Skłodowska-Curie Actions Individual Fellowship No. 101022150) and the Xunta de Galicia (Centro Singular de Investigación de Galicia, 2019-2022, grant no. ED431G2019/03).

REFERENCES

(1) Clar, E.; Stewart, D. G. Aromatic Hydrocarbons. LXV. Triangulene Derivatives1. *J. Am. Chem. Soc.* **1953**, *75* (11), 2667–2672. https://doi.org/10.1021/ja01107a035.

(2) Randić, M. Aromaticity of Polycyclic Conjugated Hydrocarbons. *Chem. Rev.* **2003**, *103* (9), 3449–3606. https://doi.org/10.1021/cr9903656.

(3) Pavliček, N.; Mistry, A.; Majzik, Z.; Moll, N.; Meyer, G.; Fox, D. J.; Gross, L. Synthesis and Characterization of Triangulene. *Nature Nanotech* **2017**, *12* (4), 308–311. https://doi.org/10.1038/nnano.2016.305.

(4) Su, J.; Telychko, M.; Song, S.; Lu, J. Triangulenes: From Precursor Design to On-Surface Synthesis and Characterization. *Angew. Chem.* **2020**, *132* (20), 7730–7740. https://doi.org/10.1002/ange.201913783.

(5) Ovchinnikov, A. A. Multiplicity of the Ground State of Large Alternant Organic Molecules with Conjugated Bonds: (Do Organic Ferromagnetics Exist?). *Theoret. Chim. Acta* **1978**, *47* (4), 297–304. https://doi.org/10.1007/BF00549259.

(6) Lieb, E. H. Two Theorems on the Hubbard Model. *Phys. Rev. Lett.* **1989**, *62* (10), 1201–1204. https://doi.org/10.1103/PhysRevLett.62.1201.

(7) Clair, S.; de Oteyza, D. G. Controlling a Chemical Coupling Reaction on a Surface: Tools and Strategies for On-Surface Synthesis. *Chem. Rev.* **2019**, *119* (7), 4717–4776. https://doi.org/10.1021/acs.chemrev.8b00601.

(8) Wang, T.; Zhu, J. Confined On-Surface Organic Synthesis: Strategies and Mechanisms. *Surface Science Reports* **2019**, *74* (2), 97–140. https://doi.org/10.1016/j.surfrep.2019.05.001.

(9) Grill, L.; Hecht, S. Covalent On-Surface Polymerization. *Nat. Chem.* **2020**, *12* (2), 115–130. https://doi.org/10.1038/s41557-019-0392-9.

(10) Cai, J.; Ruffieux, P.; Jaafar, R.; Bieri, M.; Braun, T.; Blankenburg, S.; Muoth, M.; Seitsonen, A. P.; Saleh, M.; Feng, X.; Müllen, K.; Fasel, R. Atomically Precise Bottom-up Fabrication of Graphene Nanoribbons. *Nature* **2010**, *466* (7305), 470–473. https://doi.org/10.1038/nature09211.

(11) Song, S.; Su, J.; Telychko, M.; Li, J.; Li, G.; Li, Y.; Su, C.; Wu, J.; Lu, J. On-Surface Synthesis of Graphene Nanostructures with π-Magnetism. *Chem. Soc. Rev.* **2021**, *50* (5), 3238–3262. https://doi.org/10.1039/D0CS01060J.

(12) Li, J.; Sanz, S.; Corso, M.; Choi, D. J.; Peña, D.; Frederiksen, T.; Pascual, J. I. Single Spin Localization and Manipulation in Graphene Open-Shell Nanostructures. *Nat Commun* **2019**, *10* (1), 200. https://doi.org/10.1038/s41467-018-08060-6.

(13) Mishra, S.; Beyer, D.; Eimre, K.; Kezilebieke, S.; Berger, R.; Gröning, O.; Pignedoli, C. A.; Müllen, K.; Liljeroth, P.; Ruffieux, P.; Feng, X.; Fasel, R. Topological Frustration Induces Unconventional Magnetism in a Nanographene. *Nat. Nanotechnol.* **2020**, *15* (1), 22–28. https://doi.org/10.1038/s41565-019-0577-9.

(14) Mishra, S.; Beyer, D.; Eimre, K.; Liu, J.; Berger, R.; Gröning, O.; Pignedoli, C. A.; Müllen, K.; Fasel, R.; Feng, X.; Ruffieux, P. Synthesis and Characterization of π-Extended Triangulene. *J. Am. Chem. Soc.* **2019**, *141* (27), 10621–10625. https://doi.org/10.1021/jacs.9b05319.

(15) Su, J.; Telychko, M.; Hu, P.; Macam, G.; Mutombo, P.; Zhang, H.; Bao, Y.; Cheng, F.; Huang, Z.-Q.; Qiu, Z.; Tan, S. J. R.; Lin, H.; Jelínek, P.; Chuang, F.-C.; Wu, J.; Lu, J. Atomically Precise Bottom-up Synthesis of π-Extended [5]Triangulene. *Sci. Adv.* **2019**, *5* (7), eaav7717. https://doi.org/10.1126/sciadv.aav7717.

(16) Jelínek, P. High Resolution SPM Imaging of Organic Molecules with Functionalized Tips. *J. Phys.: Condens. Matter* **2017**, *29* (34), 343002. https://doi.org/10.1088/1361-648X/aa76c7.

(17) Hieulle, J.; Castro, S.; Friedrich, N.; Vegliante, A.; Lara, F. R.; Sanz, S.; Rey, D.; Corso, M.; Frederiksen, T.; Pascual, J. I.; Peña, D. On-Surface Synthesis and Collective Spin Excitations of a Triangulene-Based Nanostar. *Angew. Chem. Int. Ed.* **2021**, anie.202108301. https://doi.org/10.1002/anie.202108301.

(18) Mishra, S.; Catarina, G.; Wu, F.; Ortiz, R.; Jacob, D.; Eimre, K.; Ma, J.; Pignedoli, C. A.; Feng, X.; Ruffieux, P.; Fernández-Rossier, J.; Fasel, R. Observation of Fractional Edge Excitations in Nanographene Spin Chains. *Nature* **2021**, *598* (7880), 287–292. https://doi.org/10.1038/s41586-021-03842-3.

(19) Mishra, S.; Beyer, D.; Eimre, K.; Ortiz, R.; Fernández-Rossier, J.; Berger, R.; Gröning, O.; Pignedoli, C. A.; Fasel, R.; Feng, X.; Ruffieux, P. Collective All-Carbon Magnetism in Triangulene Dimers**. *Angew. Chem. Int. Ed.* **2020**, *59* (29), 12041–12047. https://doi.org/10.1002/anie.202002687.

(20) Wang, X.; Sun, G.; Routh, P.; Kim, D.-H.; Huang, W.; Chen, P. Heteroatom-Doped Graphene Materials: Syntheses, Properties and Applications. *Chem. Soc. Rev.* **2014**, *43* (20), 7067–7098. https://doi.org/10.1039/C4CS00141A.

(21) Wang, X.-Y.; Yao, X.; Narita, A.; Müllen, K. Heteroatom-Doped Nanographenes with Structural Precision. *Acc. Chem. Res.* **2019**, *52* (9), 2491–2505. https://doi.org/10.1021/acs.accounts.9b00322.

(22) Wang, X.-Y.; Urgel, J. I.; Barin, G. B.; Eimre, K.; Di Giovannantonio, M.; Milani, A.; Tommasini, M.; Pignedoli, C. A.; Ruffieux, P.; Feng, X.; Fasel, R.; Müllen, K.; Narita, A. Bottom-Up Synthesis of Heteroatom-Doped Chiral Graphene Nanoribbons. *J. Am. Chem. Soc.* **2018**, *140* (29), 9104–9107. https://doi.org/10.1021/jacs.8b06210.

(23) Kawai, S.; Nakatsuka, S.; Hatakeyama, T.; Pawlak, R.; Meier, T.; Tracey, J.; Meyer, E.; Foster, A. S. Multiple

Heteroatom Substitution to Graphene Nanoribbon. *Sci. Adv.* **2018**, *4* (4), eaar7181. https://doi.org/10.1126/sciadv.aar7181.

(24) Friedrich, N.; Brandimarte, P.; Li, J.; Saito, S.; Yamaguchi, S.; Pozo, I.; Peña, D.; Frederiksen, T.; Garcia-Lekue, A.; Sánchez-Portal, D.; Pascual, J. I. Magnetism of Topological Boundary States Induced by Boron Substitution in Graphene Nanoribbons. *Phys. Rev. Lett.* **2020**, *125* (14), 146801. https://doi.org/10.1103/PhysRevLett.125.146801.

(25) Carbonell-Sanromà, E.; Hieulle, J.; Vilas-Varela, M.; Brandimarte, P.; Iraola, M.; Barragán, A.; Li, J.; Abadia, M.; Corso, M.; Sánchez-Portal, D.; Peña, D.; Pascual, J. I. Doping of Graphene Nanoribbons *via* Functional Group Edge Modification. *ACS Nano* **2017**, *11* (7), 7355–7361. https://doi.org/10.1021/acsnano.7b03522.

(26) Li, J.; Brandimarte, P.; Vilas-Varela, M.; Merino-Díez, N.; Moreno, C.; Mugarza, A.; Mollejo, J. S.; Sánchez-Portal, D.; Garcia de Oteyza, D.; Corso, M.; Garcia-Lekue, A.; Peña, D.; Pascual, J. I. Band Depopulation of Graphene Nanoribbons Induced by Chemical Gating with Amino Groups. *ACS Nano* **2020**, *14* (2), 1895–1901. https://doi.org/10.1021/acsnano.9b08162.

(27) Sandoval-Salinas, M. E.; Carreras, A.; Casanova, D. Triangular Graphene Nanofragments: Open-Shell Character and Doping. *Phys. Chem. Chem. Phys.* **2019**, *21* (18), 9069–9076. https://doi.org/10.1039/C9CP00641A.

(28) Jahn, H. A.; Bragg, W. H. Stability of Polyatomic Molecules in Degenerate Electronic States II-Spin Degeneracy. *Proceedings of the Royal Society of London. Series A - Mathematical and Physical Sciences* **1938**, *164* (916), 117–131. https://doi.org/10.1098/rspa.1938.0008.

(29) Lawrence, J.; Berdonces-Layunta, A.; Edalatmanesh, S.; Castro-Esteban, J.; Wang, T.; Mohammed, M. S. G.; Vilas-Varela, M.; Jelinek, P.; Peña4, D.; de Oteyza, D. G. Circumventing the Stability Problems of Graphene Nanoribbon Zigzag Edges. *arXiv:2107.12754 [cond-mat]* **2021**.

(30) Steiner, C.; Gebhardt, J.; Ammon, M.; Yang, Z.; Heidenreich, A.; Hammer, N.; Görling, A.; Kivala, M.; Maier, S. Hierarchical On-Surface Synthesis and Electronic Structure of Carbonyl-Functionalized One- and Two-Dimensional Covalent Nanoarchitectures. *Nat Commun* **2017**, *8* (1), 14765. https://doi.org/10.1038/ncomms14765.

(31) Field, J. E.; Venkataraman, D. HeterotriangulenesStructure and Properties. *Chem. Mater.* **2002**, *14* (3), 962–964. https://doi.org/10.1021/cm010929y.

(32) Mohammed, M. S. G.; Colazzo, L.; Robles, R.; Dorel, R.; Echavarren, A. M.; Lorente, N.; de Oteyza, D. G. Electronic Decoupling of Polyacenes from the Underlying Metal Substrate by Sp3 Carbon Atoms. *Commun Phys* **2020**, *3* (1), 159. https://doi.org/10.1038/s42005-020-00425-y.

(33) Li, J.; Sanz, S.; Castro-Esteban, J.; Vilas-Varela, M.; Friedrich, N.; Frederiksen, T.; Peña, D.; Pascual, J. I. Uncovering the Triplet Ground State of Triangular Graphene Nanoflakes Engineered with Atomic Precision on a Metal Surface. *Phys. Rev. Lett.* **2020**, *124* (17), 177201. https://doi.org/10.1103/PhysRevLett.124.177201.

(34) Di Giovannantonio, M.; Eimre, K.; Yakutovich, A. V.; Chen, Q.; Mishra, S.; Urgel, J. I.; Pignedoli, C. A.; Ruffieux, P.; Müllen, K.; Narita, A.; Fasel, R. On-Surface Synthesis of Antiaromatic and Open-Shell Indeno[2,1-*b*]Fluorene Polymers and Their Lateral Fusion into Porous Ribbons. *J. Am. Chem. Soc.* **2019**, *141* (31), 12346–12354. https://doi.org/10.1021/jacs.9b05335.

(35) Di Giovannantonio, M.; Chen, Q.; Urgel, J. I.; Ruffieux, P.; Pignedoli, C. A.; Müllen, K.; Narita, A.; Fasel, R. On-Surface Synthesis of Oligo(Indenoindene). *J. Am. Chem. Soc.* **2020**, *142* (30), 12925–12929. https://doi.org/10.1021/jacs.0c05701.

(36) Kichin, G.; Weiss, C.; Wagner, C.; Tautz, F. S.; Temirov, R. Single Molecule and Single Atom Sensors for Atomic Resolution Imaging of Chemically Complex Surfaces. *J. Am. Chem. Soc.* **2011**, *133* (42), 16847–16851. https://doi.org/10.1021/ja204624g.

(37) Stuyver, T.; Chen, B.; Zeng, T.; Geerlings, P.; De Proft, F.; Hoffmann, R. Do Diradicals Behave Like Radicals? *Chem. Rev.* **2019**, *119* (21), 11291–11351. https://doi.org/10.1021/acs.chemrev.9b00260.

(38) Berdonces-Layunta, A.; Lawrence, J.; Edalatmanesh, S.; Castro-Esteban, J.; Wang, T.; Mohammed, M. S. G.; Colazzo, L.; Peña, D.; Jelínek, P.; de Oteyza, D. G. Chemical Stability of (3,1)-Chiral Graphene Nanoribbons. *ACS Nano* **2021**, *15* (3), 5610–5617. https://doi.org/10.1021/acsnano.1c00695.

(39) Okabayashi, N.; Peronio, A.; Paulsson, M.; Arai, T.; Giessibl, F. J. Vibrations of a Molecule in an External Force Field. *Proc Natl Acad Sci USA* **2018**, *115* (18), 4571–4576. https://doi.org/10.1073/pnas.1721498115.

(40) de la Torre, B.; Švec, M.; Foti, G.; Krejčí, O.; Hapala, P.; Garcia-Lekue, A.; Frederiksen, T.; Zbořil, R.; Arnau, A.; Vázquez, H.; Jelínek, P. Submolecular Resolution by Variation of the Inelastic Electron Tunneling Spectroscopy Amplitude and Its Relation to the AFM/STM Signal. *Phys. Rev. Lett.* **2017**, *119* (16), 166001. https://doi.org/10.1103/PhysRevLett.119.166001.

(41) Kondo, J. Resistance Minimum in Dilute Magnetic Alloys. *Progress of Theoretical Physics* **1964**, *32* (1), 37–49. https://doi.org/10.1143/PTP.32.37.

(42) Frota, H. O. Shape of the Kondo Resonance. *Phys. Rev. B* **1992**, *45* (3), 1096–1099. https://doi.org/10.1103/PhysRevB.45.1096.

(43) Zhang, Y.; Kahle, S.; Herden, T.; Stroh, C.; Mayor, M.; Schlickum, U.; Ternes, M.; Wahl, P.; Kern, K. Temperature and Magnetic Field Dependence of a Kondo System in the Weak Coupling Regime. *Nat Commun* **2013**, *4* (1), 2110. https://doi.org/10.1038/ncomms3110.

(44) Ternes, M. Spin Excitations and Correlations in Scanning Tunneling Spectroscopy. *New J. Phys.* **2015**, *17* (6), 063016. https://doi.org/10.1088/1367-2630/17/6/063016.

(45) Su, X.; Li, C.; Du, Q.; Tao, K.; Wang, S.; Yu, P. Atomically Precise Synthesis and Characterization of Heptauthrene with Triplet Ground State. *Nano Lett.* **2020**, *20* (9), 6859–6864. https://doi.org/10.1021/acs.nanolett.0c02939.

(46) Merino-Díez, N.; Garcia-Lekue, A.; Carbonell-Sanromà, E.; Li, J.; Corso, M.; Colazzo, L.; Sedona, F.; Sánchez-Portal, D.; Pascual, J. I.; de Oteyza, D. G. Width-Dependent Band Gap in Armchair Graphene Nanoribbons Reveals Fermi Level Pinning on Au(111). *ACS Nano* **2017**, *11* (11), 11661–11668. https://doi.org/10.1021/acsnano.7b06765.

(47) Ketabi, N.; de Boer, T.; Karakaya, M.; Zhu, J.; Podila, R.; Rao, A. M.; Kurmaev, E. Z.; Moewes, A. Tuning the Electronic Structure of Graphene through Nitrogen Doping: Experiment and Theory. *RSC Adv.* **2016**, *6* (61), 56721–56727. https://doi.org/10.1039/C6RA07546K.

(48) Sun, Q.; Mateo, L. M.; Robles, R.; Ruffieux, P.; Lorente, N.; Bottari, G.; Torres, T.; Fasel, R. Inducing Open-Shell Character in Porphyrins through Surface-Assisted

Phenalenyl π-Extension. *J. Am. Chem. Soc.* **2020**, *142* (42), 18109–18117. https://doi.org/10.1021/jacs.0c07781.

(49) Zhao, Y.; Jiang, K.; Li, C.; Liu, Y.; Xu, C.; Zheng, W.; Guan, D.; Li, Y.; Zheng, H.; Liu, C.; Luo, W.; Jia, J.; Zhuang, X.; Wang, S. Precise Control of π-Electron Magnetism in Metal-Free Porphyrins. *J. Am. Chem. Soc.* **2020**, *142* (43), 18532–18540. https://doi.org/10.1021/jacs.0c07791.

(50) Biswas, K.; Urgel, J. I.; Xu, K.; Ma, J.; Sánchez-Grande, A.; Mutombo, P.; Gallardo-Caparrós, A.; Lauwaet, K.; Mallada, B.; de la Torre, B.; Matěj, A.; Gallego, J. M.; Miranda, R.; Jelínek, P.; Feng, X.; Écija, D. On-Surface Synthesis of a Dicationic Diazahexabenzocoronene Derivative on the Au(111) Surface. *Angewandte Chemie International Edition* **2021**, *n/a* (n/a). https://doi.org/10.1002/anie.202111863.

(51) Michaelson, H. B. The Work Function of the Elements and Its Periodicity. *J. Appl. Phys.* **1977**, *48* (11), 5.

(52) Mishra, S.; Yao, X.; Chen, Q.; Eimre, K.; Gröning, O.; Ortiz, R.; Di Giovannantonio, M.; Sancho-García, J. C.; Fernández-Rossier, J.; Pignedoli, C. A.; Müllen, K.; Ruffieux, P.; Narita, A.; Fasel, R. Large Magnetic Exchange Coupling in Rhombus-Shaped Nanographenes with Zigzag Periphery. *Nature Chemistry* **2021**, *13* (6), 581–586. https://doi.org/10.1038/s41557-021-00678-2.

(53) Rano, M.; Ghosh, S. K.; Ghosh, D. In the Quest for a Stable Triplet State in Small Polyaromatic Hydrocarbons: An *in Silico* Tool for Rational Design and Prediction. *Chem. Sci.* **2019**, *10* (40), 9270–9276. https://doi.org/10.1039/C9SC02414J.

(54) Jacob, D.; Ortiz, R.; Fernández-Rossier, J. Renormalization of Spin Excitations and Kondo Effect in Open-Shell Nanographenes. *Phys. Rev. B* **2021**, *104* (7), 075404. https://doi.org/10.1103/PhysRevB.104.075404.

# Supporting information

# Synthesis and Characterization of Nitrogen-Doped Triangulene on Metal Surfaces

Tao Wang,[1,2,‡] Alejandro Berdonces-Layunta,[1,2,‡] Niklas Friedrich,[3] Manuel Vilas-Varela,[4] Jan Patrick Calupitan,[2,*] Jose Ignacio Pascual,[3,5] Diego Peña,[4] David Casanova,[1,5] Martina Corso,[1,2] and Dimas G. de Oteyza[1,2,5,*]

[1]Donostia International Physics Center, 20018 San Sebastián, Spain

[2]Centro de Fisica de Materiales CFM/MPC, CSIC-UPV/EHU, 20018 San Sebastián, Spain

[3]CIC nanoGUNE BRTA, 20018 San Sebastián, Spain

[4]Centro Singular de Investigación en Química Biolóxica e Materiais Moleculares (CiQUS) and Departamento de Química Orgánica, Universidade de Santiago de Compostela, 15782 Santiago de Compostela, Spain

[5]Ikerbasque, Basque Foundation for Science, 48009 Bilbao, Spain

# Methods

## STM Experiments

STM measurements were performed using a commercial Scienta-Omicron LT-STM at 4.3 K. The system consists of a preparation chamber with a typical pressure in the low $10^{-10}$ mbar regime and a STM chamber with a pressure in the $10^{-11}$ mbar range. The Au(111) and Ag(111) crystals were cleaned *via* two cycles of $Ar^+$ sputtering and annealing (720 K for Au and 700 K for Ag). Ketone substituted triangulene precursor molecule was evaporated from a home-built evaporator at 460 K. Hydrogenation of the sample was achieved with a hydrogen cracking source with a leak valve. The preparation chamber was first filled to a pressure of $2 \times 10^{-7}$ mbar, after which the tungsten tube was heated to around 2800 K with a heating power of 80 W. The sample was then placed in front of the source for 2 minutes.

All STM and STS measurements were performed at 4.3 K. To obtain BR-STM images, the tip was functionalized with a CO molecule that was picked up from the metal surfaces. CO was deposited onto the sample *via* a leak valve at a pressure of approximately $5 \times 10^{-9}$ mbar and a maximum sample temperature of 7.0 K. CO can be picked up with a metallic tip by scanning with a high current and negative bias (*e.g.* I=1 nA, U=−0.5 V). Functionalization of the tip with a Cl atom is achieved by approaching the tip by 350 pm from initial stabilization conditions of 100 pA and 100 mV with the feedback off, on the top of the deposited NaCl island on Au(111). After pickup, an increase in resolution is seen and a vacancy in the NaCl island can also be observed, which was reported in detail in our previous work.[1] dI/dV measurements were recorded with the internal lock-in of the system. The oscillation frequency used in experiments is 797 Hz. The amplitude for each spectrum is shown in Figure captions.

## DFT calculations

DFT calculations were performed by the Gaussian 16 package[2] using the M06-2X functional and 6-311G(d,p) basis set. Results were visualized by using the software Gaussview[3] and the squares of the wavefunction were generated using the cubman module.

For the positively-charged species, the atomic positions of the neutral N-doped triangulene[4] were used as input for the optimization. Initial optimization set to charge=+1 and spin=1 yielded a $D_{3h}$ geometry, independent of whether the starting geometries corresponded to $D_{3h}$ or $C_{2v}$ geometry. Similarly, starting from any of the neutral $C_{2v}$ or $D_{3h}$ geometries led to a $D_{3h}$ geometry for the negatively-charged species. All geometries and symmetries were confirmed further by re-performing calculations with tight convergence criteria using the keywords *opt=tight int=ultrafine*. Ionization energies (Figure S9) were obtained as the electronic energy difference between potential energy surface minima of neutral and cationic species.

## Solution synthesis

Starting materials were purchased from TCI and Sigma-Aldrich and used without further purification. O,O',O''-Amino-trisbenzoic acid-trimethylester (**S3**) and 4H-benzo[9,1]quinolizino[3,4,5,6,7-defg]acridine-4,8,12-trione (**1**) were synthesized following slightly modified literature procedures.[5,6] Reactions were carried out in flame-dried glassware and under an inert atmosphere (Ar) using Schlenk techniques. Thin-layer chromatography (TLC) was performed on Silica Gel 60 F-254 plates (Merck). Column chromatography was performed on silica gel (40-60 μm). NMR spectra were recorded on a Varian Mercury 300 spectrometer.

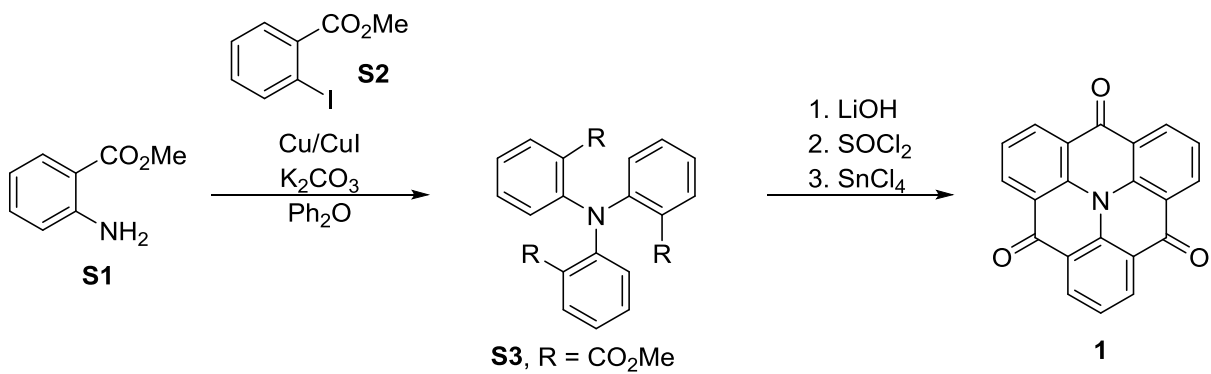

**Figure S1.** Synthesis of compound **1**.

**Synthesis of compound S3.** A mixture of methyl anthranilate (**S1**) (2.0 ml, 15.5 mmol), methyl 2-iodobenoate (**S2**) (6.5 ml, 43.3 mmol), $K_2CO_3$ (5.3 g, 38.8 mmol), Cu (0.2 g, 3.1 mmol) and CuI (0.3 g, 1.6 mmol) in diphenylether (15 ml) was heated at 190 °C under Ar for 72 hours. The solvent was removed under reduced pressure and the residue was purified by column chromatography ($SiO_2$; 4:1 hexane/ethyl acetate) to afford **S3** (3.6 g, 55%) as a yellow solid. $^1$H NMR ($CDCl_3$) δ 7.60 (m, 3H), 7.36 (m, 3H), 7.07 (m, 6H), 3.37 (s, 9H) ppm.

**Synthesis of compound 1.** A mixture of **S3** (0.50 g, 1.2 mmol) and $LiOH·H_2O$ (0.90 g, 21 mmol) in THF: $H_2O$ (4:1, 25 mL) was refluxed for 6 h. After cooling to room temperature, the mixture was diluted with ethyl acetate (10 mL) and water (10 mL), and the phases were separated. The aqueous phase was acidified with concentrated HCl until pH = 2 and then extracted with ethyl acetate (3x10 mL). Organic extracts were combined, dried over anhydrous $Na_2SO_4$, filtered and evaporated. The crude product was dissolved in dichloromethane (20 mL) and two drops of DMF were added, followed by $SOCl_2$ (1.7 mL). The resulting mixture was refluxed for 3 h and then cooled to 0 °C. Then, $SnCl_4$ (1.7 mL) was added dropwise. The resulting mixture was refluxed for 16 h and the formed precipitate was collected by filtration and washed with methanol (2x20 mL). The obtained solid was then suspended in 1 M NaOH (20 mL) and stirred for 30 min, filtered, and washed with water (3x20 mL), methanol (2x20 mL) and acetone (2x20 mL) to afford **1** (115 mg, 45%) as a grey solid. $^1$H-NMR (10% TFA in $CDCl_3$) δ 9.19 (d, J = 7.7 Hz, 6H), 8.05 (t, J = 7.8 Hz, 3H). ppm.

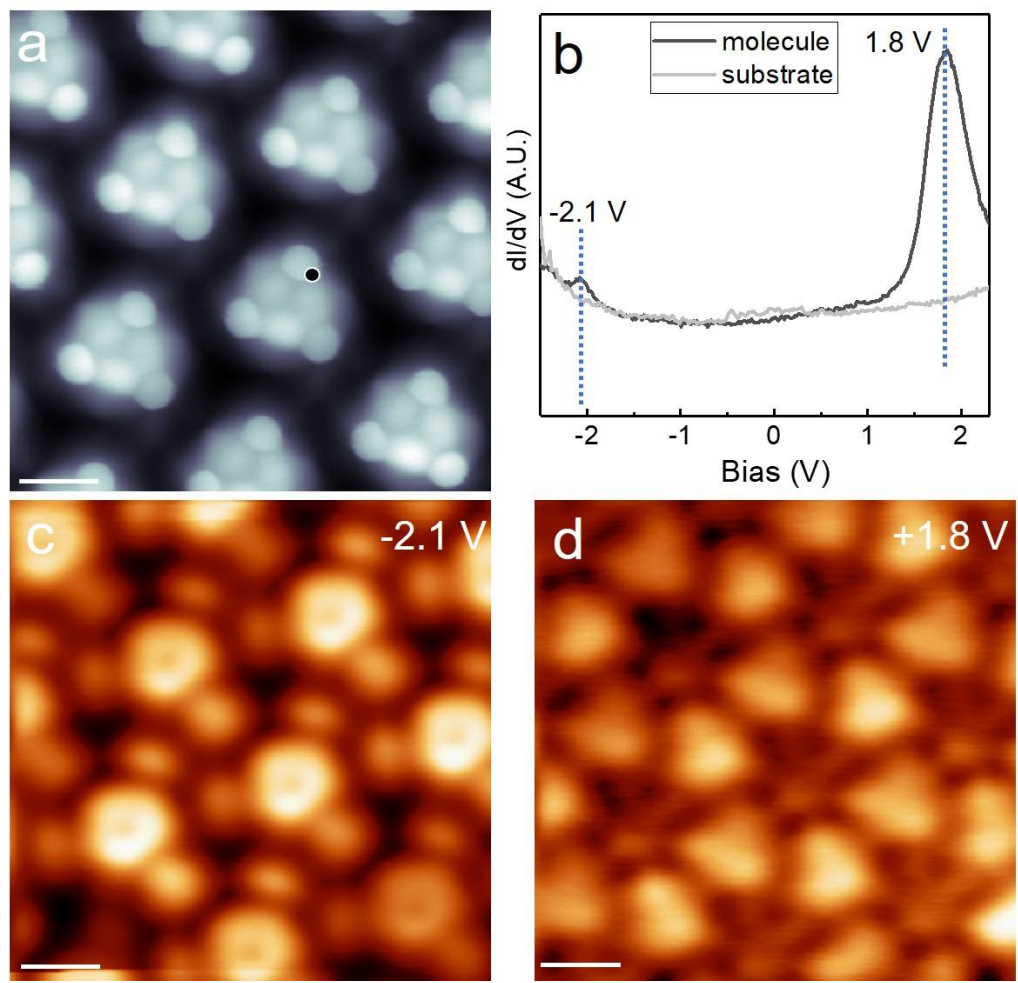

**Figure S2**. (a) BR-STM image of ketone substituted N-triangulenes on Au(111). (b) dI/dV spectra taken on N-triangulene and the Au(111) substrate with a CO-functionalized probe, respectively. The dI/dV spectroscopy on ketone substituted N-triangulene (the position is marked in (a)) presents two prominent peaks, at −2.1 and 1.8 V. (c,d) dI/dV maps taken at −2.1 and 1.8 V. All the scale bar are 5 Å.

dI/dV maps at −2.1 and 1.8 V should be associated with the HOMO and LUMO of ketone substituted N-triangulene respectively, in agreement with previous work.[7]

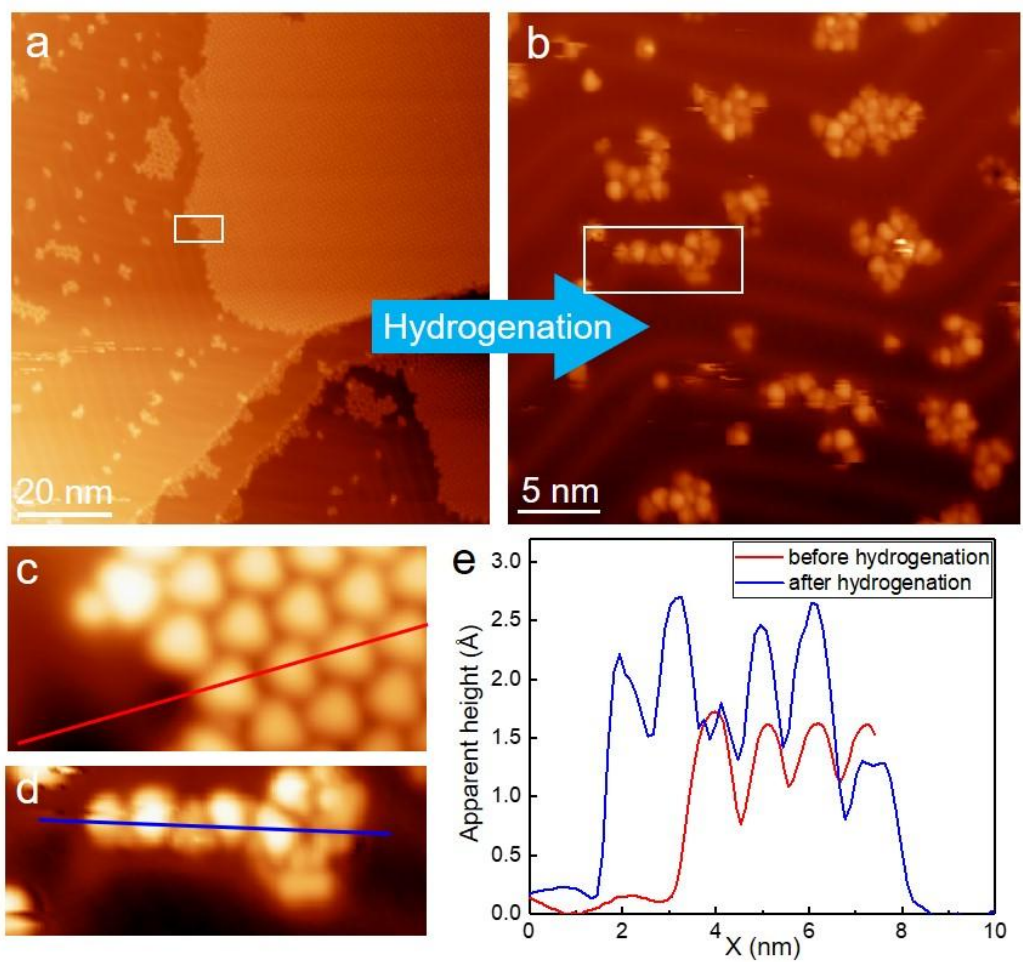

**Figure S3.** (a) large-area STM image of the ketone substituted N-triangulene covered sample. (b) STM image recorded after H reduction of (a). (c,d) Zoom-in STM image of the white framed region in (a) and (b) respectively. (e) Relative apparent height profiles of the lines shown in (c) and (d). The hydrogenated molecules are much higher than molecules before hydrogenation (~2.7 Å *vs.*~1.6 Å), indicating the existence of *sp*$^3$ carbons after hydrogenation. Scanning parameters: (a) U=1 V, I= 100 pA; (b-d) U=−1 V, I=−100 pA;

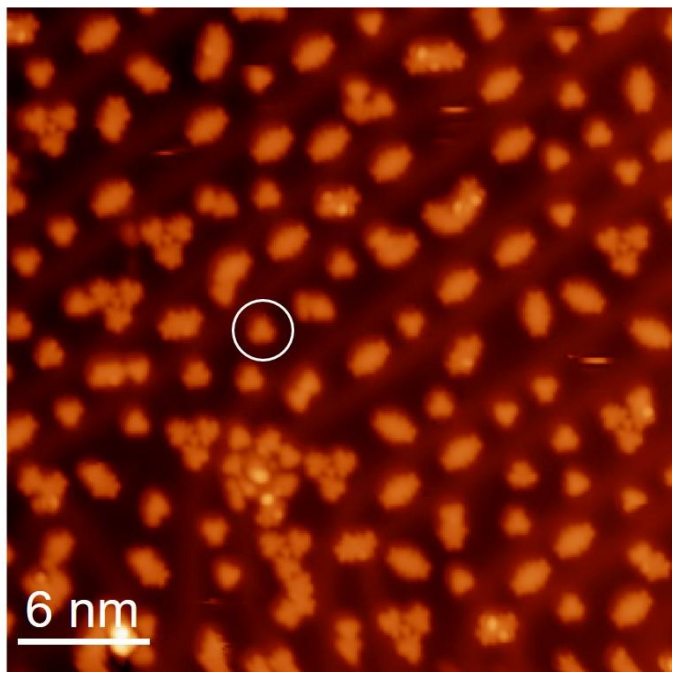

**Figure S4**. STM image of the sample prepared by annealing a hydrogenated sample to 300 ºC. Almost all the molecules (>95%) are planarized at these conditions. The target product N-triangulene is directly obtained, as exampled by the white circle marked one. Scanning parameters: (a) U=−1 V, I=−100 pA.

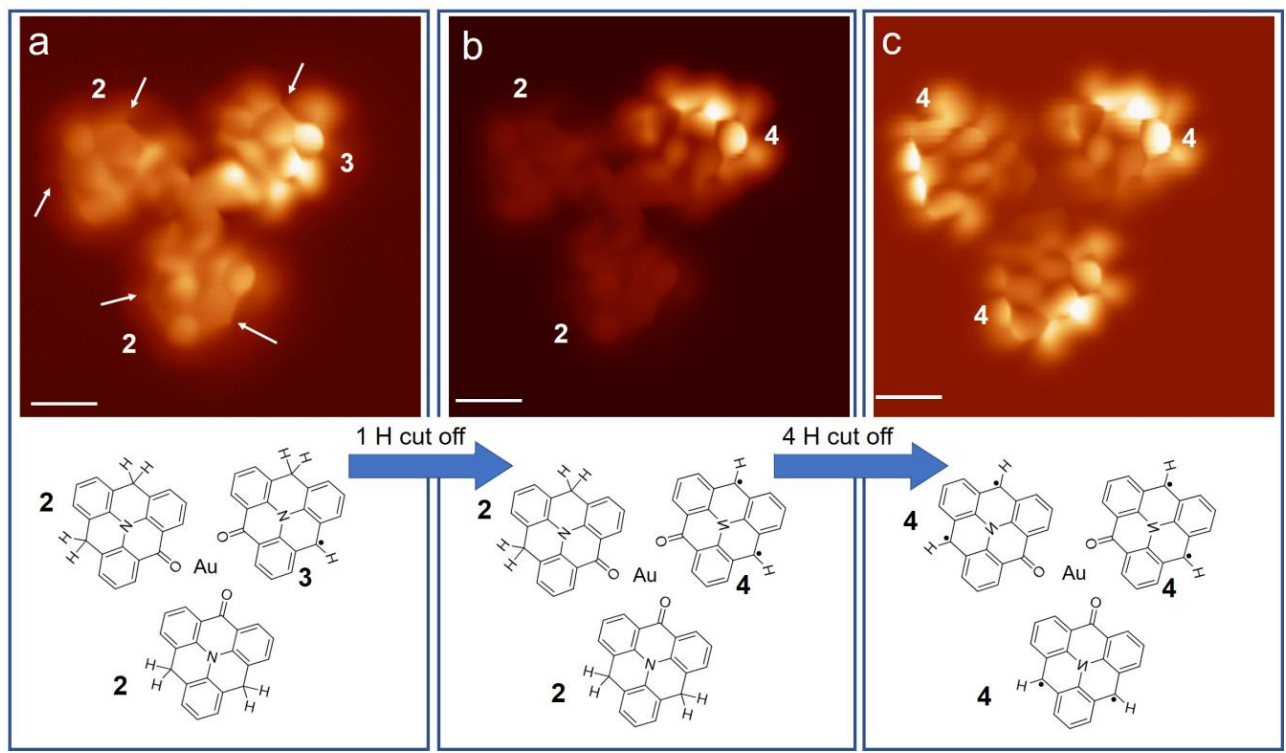

**Figure S5**. Tip manipulations on a trimer composed of two **2** and one **3**. (a) BR-STM image of the trimer structure. One monomer has one additional hydrogen while other two have two additional hydrogens, as illustrated in the chemical structures below. The rings with $sp^3$ carbons are obviously much larger than other rings, as pointed by white arrows. From (a) to (b), we cut off the additional hydrogens on **3**, producing **4**. From (b) to (c), we cut off all four remaining hydrogens and all the three monomers correspond to molecule **4**. Note that middle oxygen cannot be removed by the conventional tip manipulations. Scanning parameters: (a-c) U=5 mV. All the scale bars are 5 Å.

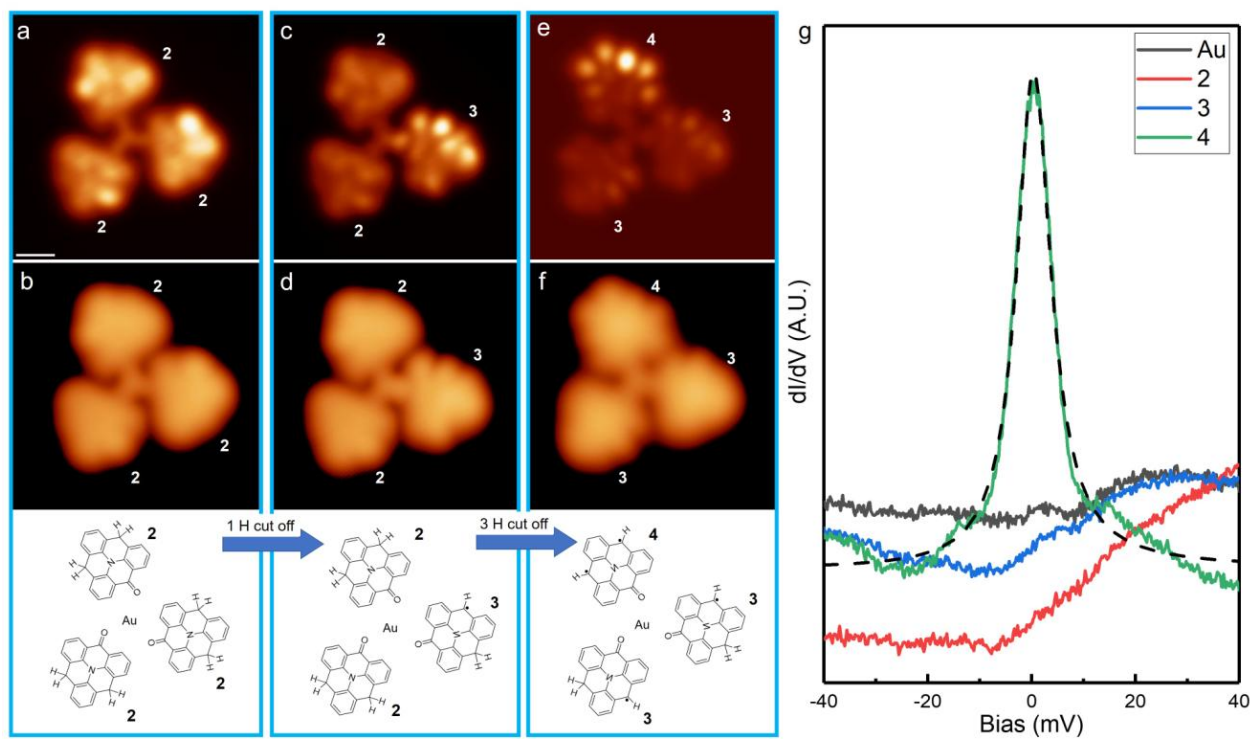

**Figure S6**. Tip manipulations on another trimer composed of three **2** to generate **3** and **4**. A Cl tip was used in all these processes. Molecular chemical structures are illustrated below the corresponding STM images. (a,b) Constant-height and constant-current STM images, showing the trimer composed of three **2**. The scale bar in (a) is 5 Å. (b,c) Constant-height and constant-current STM images taken after one H was removed (producing **3**). An obvious difference between (a) and (c), or (b) and (d) can be recognized, though without BR-STM image. (e,f) Constant-height and constant-current STM images taken after further tip manipulations which obtained two **3** and one **4**. The difference between **3** and **4** is also clearly visible. (f) dI/dV spectra taken on Au substrate, **2**, **3**, and **4**, respectively. The Kondo resonance is only observed on **4**. It is fitted by a Frota function (black dotted line) and a FWHM of 8.12±0.10 mV is obtained, in excellent agreement with the value in Figure 2 (8.1±0.6 mV). Scanning parameters: (a,c,e) U=5 mV; (b,d) U=40 mV, I=100 pA; (f) U=1 V, I =100 pA.

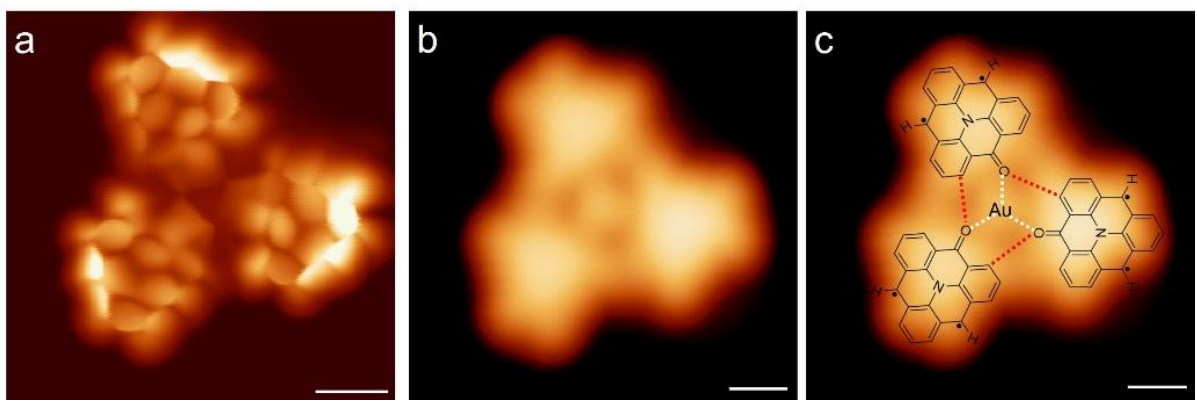

**Figure S7**. High-resolution STM images showing the Au adatom inside the trimer structure. (a) BR-STM image of a trimer structure composed of **4**. (b) shows the constant-current STM image of (a), where the middle Au adatom is clearly visible. The corresponding chemical structure matches the STM image in (b), as revealed in (c). White and red dotted lines present O···Au coordination interactions and hydrogen bonds respectively. Scanning parameters: (a) U=5 mV; (b,c) U=−0.5 V, I=−100 pA. All the scale bars are 5 Å.

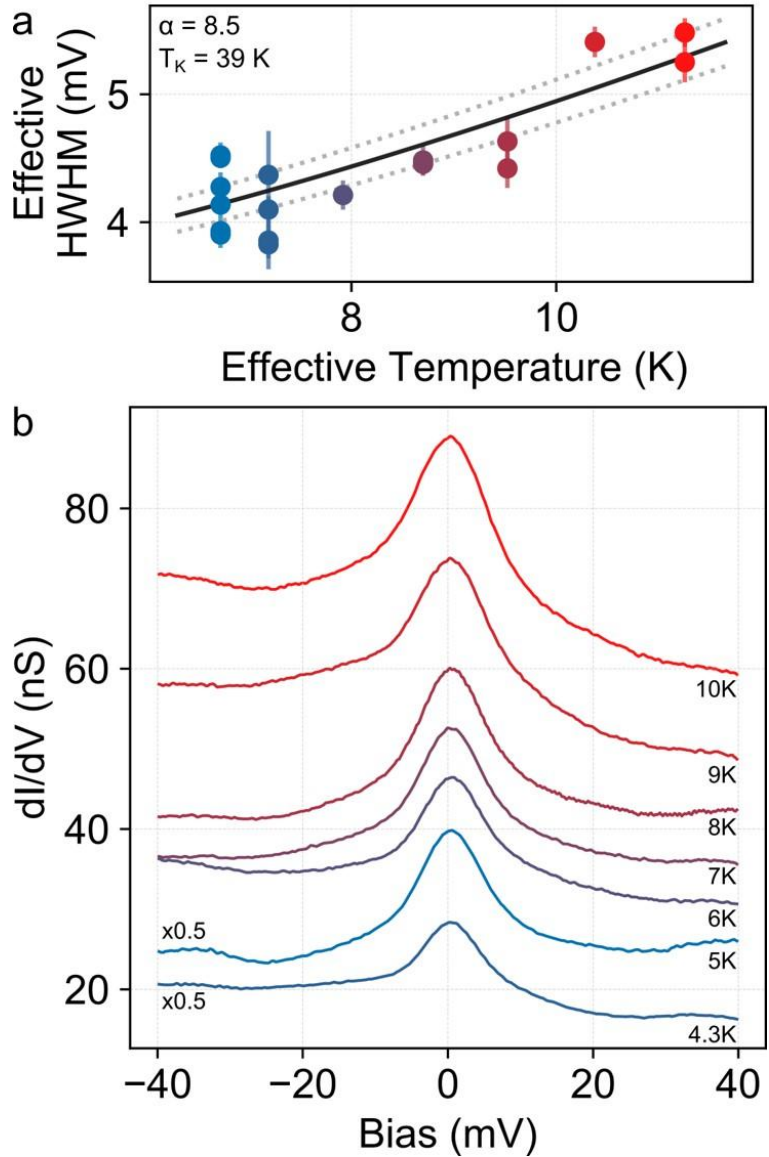

**Figure S8**. (a) Fermi liquid model of the Kondo resonance of molecule **6** on Au(111). (b) Example raw data at the different temperatures used for (a). Following Mishra et al.,[8] we extract the linewidth of the Kondo resonance by fitting a Frota function to the experimental data and correct the extracted half width at half maximum (HWHM) to an effective HWHM$_{eff}$ to account for finite temperature of the tip[9] at each temperature point. An effective temperature $T_{eff}$ is used to account for the modulation of the lock-in amplifier.[10] Finally, we extract the Kondo temperature *via* the empirical found formula HWHM$_{eff}$ = $\frac{1}{2}\sqrt{(\alpha k_B T_{eff})^2 + (2k_B T_K)^2}$ ,[11] with the Boltzmann constant k$_B$ and the fitting parameters α and T$_K$. The values α= 8.5±0.3 and T$_K$ = 39±1 K are obtained.

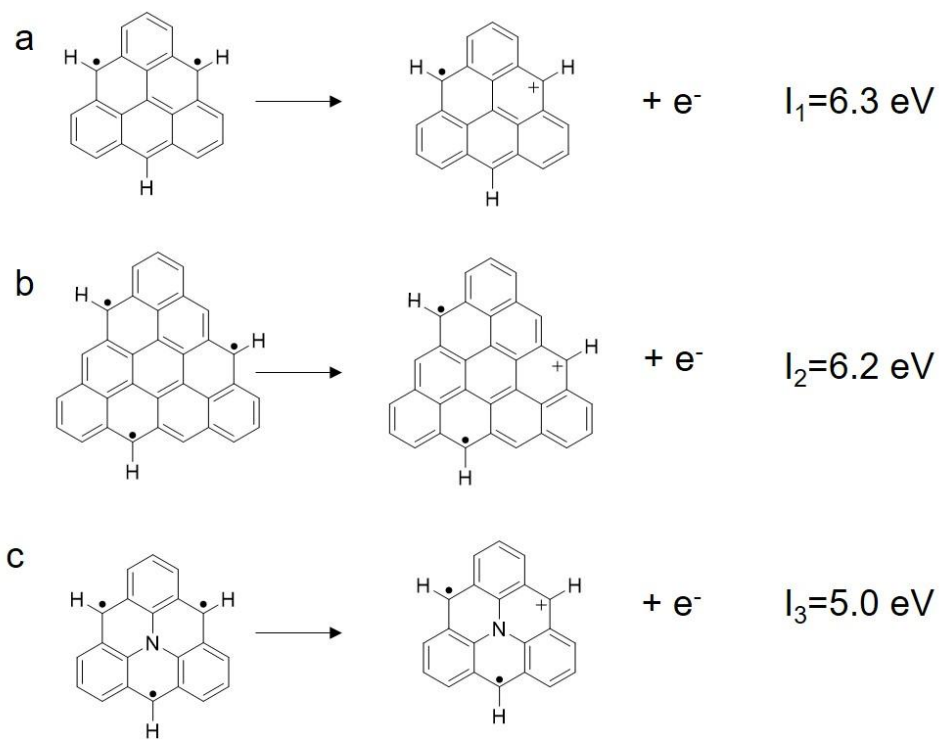

**Figure S9**. The ionization energies of (a) triangulene, (b) [4]triangulene, and (c) N-triangulene computed at the M06-2X/6-311G(d,p) level.

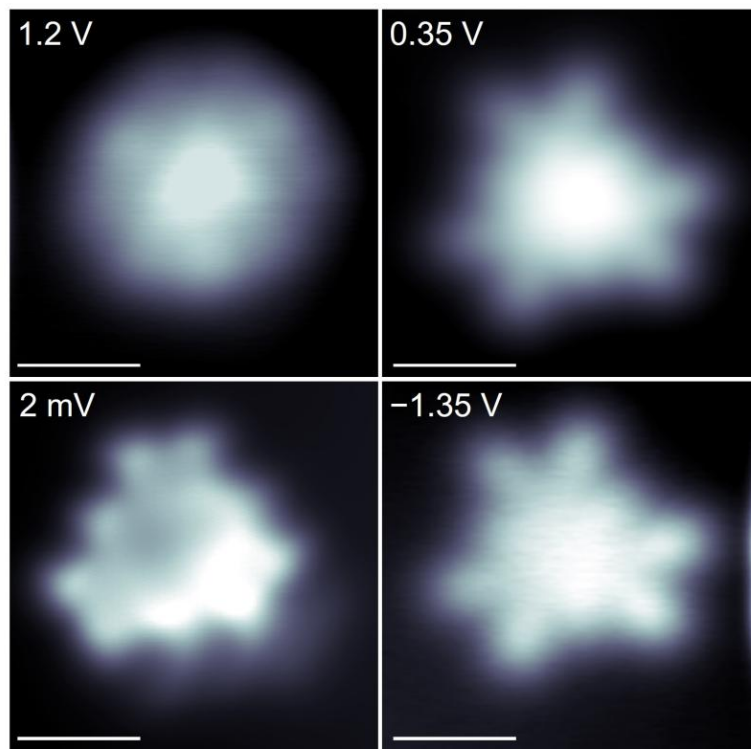

**Figure S10**. Constant-height dI/dV maps corresponding to LUMO, SUMOs, SOMOs and Kondo resonance (at 2 mV) of N-triangulene on Au(111) using a Cl-functionalized probe. All the scale bars are 5 Å.

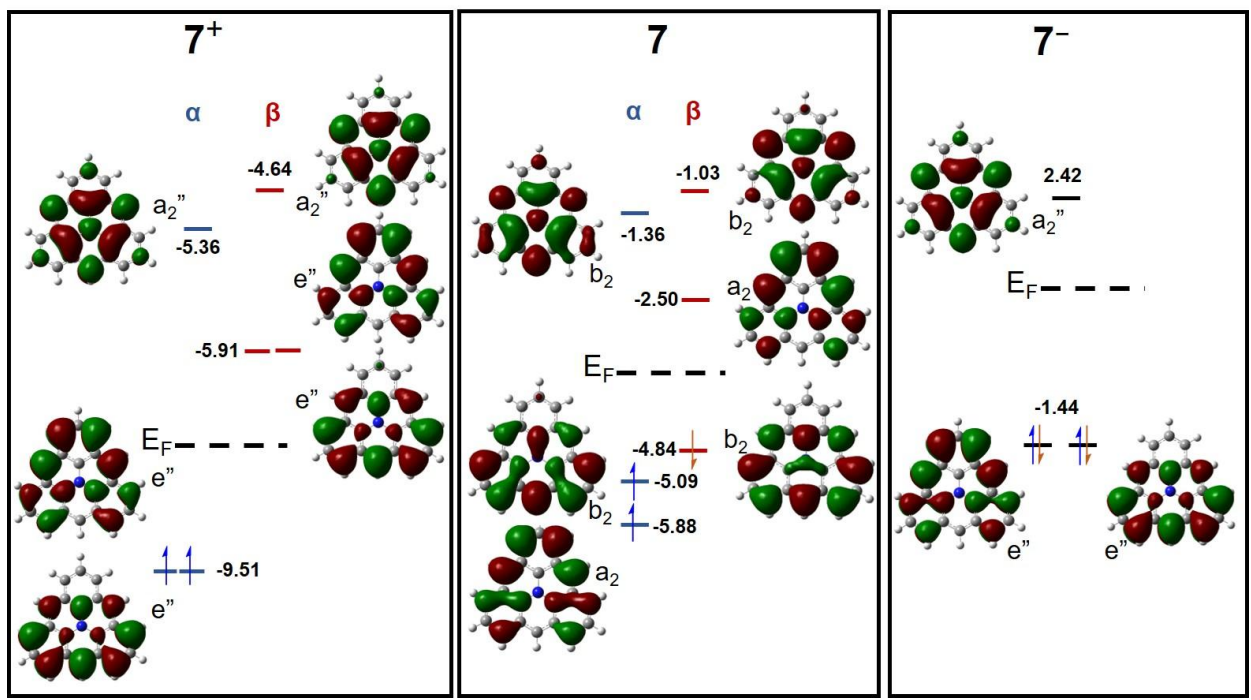

**Figure S11**. DFT calculated wave functions and energy levels of the positively charged, neutral, and negatively charged N-triangulenes, along with their corresponding energies and orbital symmetry.

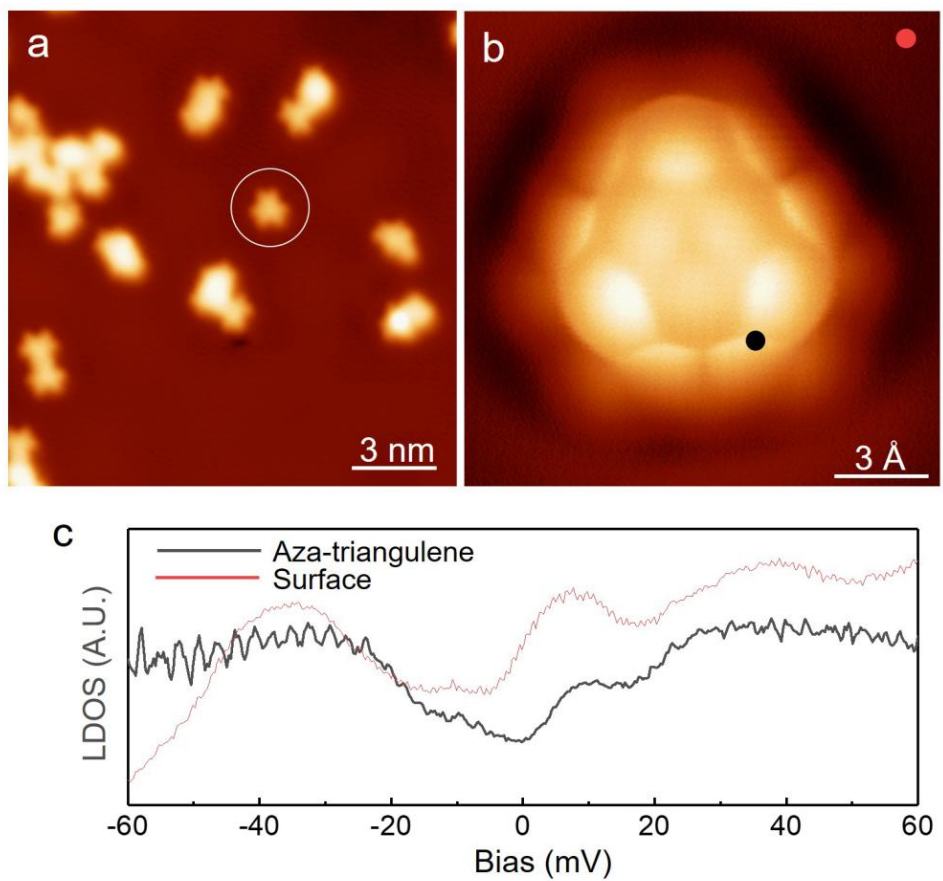

**Figure S12**. (a) STM image showing different molecular products obtained by annealing the hydrogenated ketone-N-triangulene at 300 °C on the Ag(111) sample. N-triangulene as marked by white circle coexist with some polymers generated from the fusion of N-triangulenes. Scanning parameters: U=−0.8 V; I=−80 pA. (b) Constant-height BR-STM image of N-triangulene on Ag(111) at 5 mV, using a CO-functionalized tip. (c) Low-energy dI/dV spectra taken on N-triangulene and the bare Ag(111) substrate (the positions are marked in (b)), respectively.

# References

(1) Lawrence, J.; Brandimarte, P.; Berdonces-Layunta, A.; Mohammed, M. S. G.; Grewal, A.; Leon, C. C.; Sánchez-Portal, D.; de Oteyza, D. G. Probing the Magnetism of Topological End States in 5-Armchair Graphene Nanoribbons. *ACS Nano* **2020**, *14* (4), 4499–4508. https://doi.org/10.1021/acsnano.9b10191.

(2) *Gaussian 16, Revision C.01, Frisch, M. J.; Trucks, G. W.; Schlegel, H. B.; Scuseria, G. E.; Robb, M. A.; Cheeseman, J. R.; Scalmani, G.; Barone, V.; Petersson, G. A.; Nakatsuji, H.; Li, X.; Caricato, M.; Marenich, A. V.; Bloino, J.; Janesko, B. G.; Gomperts, R.; Mennucci, B.; Hratchian, H. P.; Ortiz, J. V.; Izmaylov, A. F.; Sonnenberg, J. L.; Williams-Young, D.; Ding, F.; Lipparini, F.; Egidi, F.; Goings, J.; Peng, B.; Petrone, A.; Henderson, T.; Ranasinghe, D.; Zakrzewski, V. G.; Gao, J.; Rega, N.; Zheng, G.; Liang, W.; Hada, M.; Ehara, M.; Toyota, K.; Fukuda, R.; Hasegawa, J.; Ishida, M.; Nakajima, T.; Honda, Y.; Kitao, O.; Nakai, H.; Vreven, T.; Throssell, K.; Montgomery, J. A., Jr.; Peralta, J. E.; Ogliaro, F.; Bearpark, M. J.; Heyd, J. J.; Brothers, E. N.; Kudin, K. N.; Staroverov, V. N.; Keith, T. A.; Kobayashi, R.; Normand, J.; Raghavachari, K.; Rendell, A. P.; Burant, J. C.; Iyengar, S. S.; Tomasi, J.; Cossi, M.; Millam, J. M.; Klene, M.; Adamo, C.; Cammi, R.; Ochterski, J. W.; Martin, R. L.; Morokuma, K.; Farkas, O.; Foresman, J. B.; Fox, D. J. Gaussian, Inc., Wallingford CT, 2016.*

(3) *GaussView, Version 6, Dennington, Roy; Keith, Todd A.; Millam, John M. Semichem Inc., Shawnee Mission, KS, 2016.*

(4) Sandoval-Salinas, M. E.; Carreras, A.; Casanova, D. Triangular Graphene Nanofragments: Open-Shell Character and Doping. *Phys. Chem. Chem. Phys.* **2019**, *21* (18), 9069–9076. https://doi.org/10.1039/C9CP00641A.

(5) Fang, Z.; Teo, T.-L.; Cai, L.; Lai, Y.-H.; Samoc, A.; Samoc, M. Bridged Triphenylamine-Based Dendrimers: Tuning Enhanced Two-Photon Absorption Performance with Locked Molecular Planarity. *Org. Lett.* **2009**, *11* (1), 1–4. https://doi.org/10.1021/ol801238n.

(6) Field, J. E.; Venkataraman, D. Heterotriangulenes─Structure and Properties. *Chem. Mater.* **2002**, *14* (3), 962–964. https://doi.org/10.1021/cm010929y.

(7) Steiner, C.; Gebhardt, J.; Ammon, M.; Yang, Z.; Heidenreich, A.; Hammer, N.; Görling, A.; Kivala, M.; Maier, S. Hierarchical On-Surface Synthesis and Electronic Structure of Carbonyl-Functionalized One- and Two-Dimensional Covalent Nanoarchitectures. *Nat Commun* **2017**, *8* (1), 14765. https://doi.org/10.1038/ncomms14765.

(8) Mishra, S.; Beyer, D.; Eimre, K.; Kezilebieke, S.; Berger, R.; Gröning, O.; Pignedoli, C. A.; Müllen, K.; Liljeroth, P.; Ruffieux, P.; Feng, X.; Fasel, R. Topological Frustration Induces Unconventional Magnetism in a Nanographene. *Nat. Nanotechnol.* **2020**, *15* (1), 22–28. https://doi.org/10.1038/s41565-019-0577-9.

(9) Zhang, Y.; Kahle, S.; Herden, T.; Stroh, C.; Mayor, M.; Schlickum, U.; Ternes, M.; Wahl, P.; Kern, K. Temperature and Magnetic Field Dependence of a Kondo System in the Weak Coupling Regime. *Nat Commun* **2013**, *4* (1), 2110. https://doi.org/10.1038/ncomms3110.

(10) Girovsky, J.; Nowakowski, J.; Ali, Md. E.; Baljozovic, M.; Rossmann, H. R.; Nijs, T.; Aeby, E. A.; Nowakowska, S.; Siewert, D.; Srivastava, G.; Wäckerlin, C.; Dreiser, J.; Decurtins, S.; Liu, S.-X.; Oppeneer, P. M.; Jung, T. A.; Ballav, N. Long-Range Ferrimagnetic Order in a Two-Dimensional Supramolecular Kondo Lattice. *Nature Communications* **2017**, *8* (1), 15388. https://doi.org/10.1038/ncomms15388.

(11) Nagaoka, K.; Jamneala, T.; Grobis, M.; Crommie, M. F. Temperature Dependence of a Single Kondo Impurity. *Phys. Rev. Lett.* **2002**, *88* (7), 077205. https://doi.org/10.1103/PhysRevLett.88.077205.